\documentclass[submission, Proceedings]{SciPost}

\hypersetup{
    colorlinks,
    linkcolor={red!50!black},
    citecolor={blue!50!black},
    urlcolor={blue!80!black}
}

\usepackage[bitstream-charter]{mathdesign}
\DeclareSymbolFont{usualmathcal}{OMS}{cmsy}{m}{n}
\DeclareSymbolFontAlphabet{\mathcal}{usualmathcal}

\begin{document}

\begin{center}{\Large \textbf{
Energy loss due to defect creation in solid state detectors
}}\end{center}

\begin{center}
Matti Heikinheimo\textsuperscript{1$\star$},
Sebastian Sassi\textsuperscript{1},
Kimmo Tuominen\textsuperscript{1},
Kai Nordlund\textsuperscript{1} and
Nader Mirabolfathi\textsuperscript{2}
\end{center}

\begin{center}
{\bf 1} Helsinki Institute of Physics and Department of Physics, University of Helsinki,\\
P.O.Box 64, FI-00014 University of Helsinki, Finland
\\
{\bf 2} Department of Physics and Astronomy and the Mitchell Institute for Fundamental Physics and Astronomy,
Texas A\&M University, College Station, TX 77843, USA
\\
* matti.heikinheimo@helsinki.fi
\end{center}

\begin{center}
\today
\end{center}


\definecolor{palegray}{gray}{0.95}
\begin{center}
\colorbox{palegray}{
  \begin{tabular}{rr}
  \begin{minipage}{0.1\textwidth}
    \includegraphics[width=30mm]{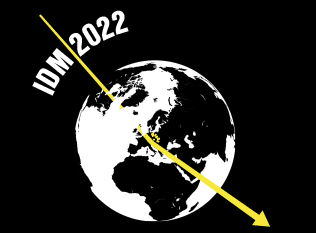}
  \end{minipage}
  &
  \begin{minipage}{0.85\textwidth}
    \begin{center}
    {\it 14th International Conference on Identification of Dark Matter}\\
    {\it Vienna, Austria, 18-22 July 2022} \\
    \doi{10.21468/SciPostPhysProc.?}\\
    \end{center}
  \end{minipage}
\end{tabular}
}
\end{center}

\section*{Abstract}
{\bf
The threshold displacement energy in solid state detector materials varies from several eV to $\lesssim\! 100$ eV. If a stable or long lived defect is created as a result of a nuclear recoil event, some part of the recoil energy is stored in the deformed lattice and is therefore not observable in a phonon detector. Thus, an accurate model of this effect is necessary for precise calibration of the recoil energy measurement in low threshold phonon detectors.
Furthermore, the sharpness of the defect creation threshold varies between materials. For a hard material such as diamond, the sharp threshold will cause a sudden onset of the energy loss effect, resulting in a prominent peak in the observed recoil spectrum just below the threshold displacement energy. We describe how this effect can be used to discriminate between nuclear and electron recoils using just the measured recoil spectrum.
}


\section{Introduction}
\label{sec:intro}
Direct detection of low mass dark matter requires detectors sensitive to recoil energies well below the keV scale. Phonon-mediated detectors with ${\cal O}$(eV) resolution are the most appropriate detectors for this purpose. Phonons are among the lowest energy quantum excitations (compred to \textit{e.g.} ionization and scintillation) that can be detected after particle interactions. In addition to their excellent signal to noise, phonon-mediated detectors offer an interaction-type independent (nuclear or electron recoil) energy measurement. Many groups have recently achieved energy resolution within the ${\cal O}$(eV) scale \cite{Strauss:2017cam, CPD:2020xvi,Verma:2022tkq}. 

Recently several experiments \cite{CRESST:2019axx,CRESST:2019jnq,DAMIC:2020cut,EDELWEISS:2019vjv,EDELWEISS:2020fxc,CRESST:2017ues,NUCLEUS:2019kxv,SENSEI:2020dpa,SuperCDMS:2018mne,SuperCDMS:2020ymb} have observed a steeply rising event rate at low energies, $E_{\rm r} \lesssim 1000$ eV. The origin of these events is currently unknown, and understanding their physical character is a question of great interest for both the DM and coherent neutrino scattering experiments~\cite{Proceedings:2022hmu}. Most of the anticipated background sources, such as photons or electrons, would give rise to electron recoils. Therefore the identification of the nuclear/electron recoil character of these events would add an important piece of information towards understanding and mitigating this background. 

Nuclear recoils at these energies can result in lattice defects, while electron recoils are not expected to do so. The formation of defects stores a part of the recoil energy in the crystal, resulting in a quenching of the measured energy in phonons. At higher recoil energy this quenching factor becomes approximately constant and can be absorbed in the calibration of the energy scale, but for recoil energies close to the threshold displacement energy the effect can be highly nonlinear, affecting not just the overall energy calibration but also the shape of the measured recoil spectrum \cite{Kadribasic:2020pwx,Sassi:2022njl}. In diamond, the energy loss effect turns on sharply at $\gtrsim 30$ eV. The affected recoil events are shifted towards lower enegy, resulting in a peak in the observed spectrum around $\sim\! 30$ eV, and a dip around $\sim\! 60$ eV, compared to the underlying true recoil spectrum. Gram scale diamond based detectors are expected to offer a resolution that is superior to the existing technologies~\cite{Kurinsky:2019pgb,Abdelhameed:2022skh}. In the following we will describe how the energy loss induced peak in the nuclear recoil spectrum in a diamond detector can be utilized to test if the observed spectrum originates from nuclear recoils.

\section{Energy loss}

\begin{figure}[tb]
\centering
\includegraphics[width=0.5\textwidth]{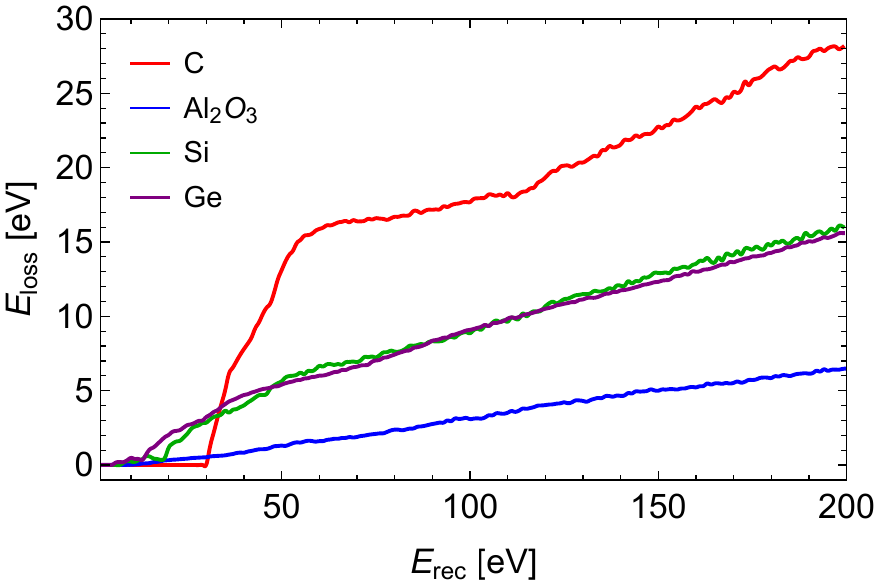}
\caption{Energy loss, averaged over recoil direction, as a function of the recoil energy in silicon, germanium, sapphire and diamond. The figure has appeared before in~\cite{Heikinheimo:2021syx}.}
\label{ElossAverage}
\end{figure}

The energy stored in the crystal lattice as a function of the nuclear recoil direction and energy were obtained from molecular dynamics (MD) simulations, described in~\cite{Sassi:2022njl}. We have simulated multiple materials, for which the numerical results can be obtained from online repository\footnote{\url{https://github.com/sebsassi/elosssim}}. The energy loss averaged over the recoil direction is shown in figure \ref{ElossAverage} for four detector materials, namely silicon (Si), germanium (Ge), sapphire (${\rm Al_2O_3}$) and diamond (C). The sharp, nonlinear threshold in diamond is evident in the figure, while sapphire shows almost linear behaviour throughout the energy range. Silicon and germanium feature nonlinear threshold behaviour, but at lower energies compared to diamond, and with a much smaller step in the stored energy. Therefore, the resulting effect on the observed spectrum should be largest for diamond, and almost negligible for the other materials.

To obtain the measured recoil spectrum in a phonon measurement after the energy loss for a given physical nuclear recoil spectrum, we perform a simulation as follows: first, the recoil direction surface (unit sphere) is divided into solid angle bins corresponding to the directions for which the MD simulations were performed as described in~\cite{Sassi:2022njl}. Each solid angle bin is then further divided into energy bins using 1~eV intervals, for which we obtain the event rate from the given physical spectrum. This procedure results in a binned underlying event rate, reflecting the energy deposited by nuclear scattering events. We then sample this underlying event rate, and for each sampled event with recoil energy $E_{\rm r}$ and recoil direction $\hat{q}$ we find the observed energy as
\begin{equation}
E_{\rm obs} = E_{\rm r} - E_{\rm loss}(E_{\rm r},\hat{q}) + E_\sigma,
\label{eq:eloss}
\end{equation}
where $E_{\rm loss}(E_{\rm r},\hat{q})$ is obtained from the MD simulations and $E_\sigma$ is a random number drawn from a Gaussian distribution with zero mean and standard deviation $\sigma$ representing the energy resolution of the detector, for which we use $\sigma = 1$~eV. The result after this sampling is the binned observed recoil rate, which we then sum over the solid angle bins as the detector is not capable of observing the recoil direction, to obtain the observed recoil spectrum $dR/dE_{\rm obs}$.

\section{Identifying nuclear recoils}
To show how the energy loss effect can be utilized to confirm that the observed spectrum originates from nuclear recoils, we use as a template a three component fit to the low energy excess spectrum, of the form
\begin{equation}
f(x) = Ae^{-\alpha x}+Bx^\beta+C,
\label{nucleusfit}
\end{equation}
where $x= E_{\rm r}/{\rm eV}$. We fit this template to three data sets from Nucleus 1g prototype \cite{NUCLEUS:2019kxv}, SuperCDMS-CPD \cite{SuperCDMS:2020aus} and Edelweiss \cite{EDELWEISS:2019vjv}. The best fit parameters are shown in table \ref{bestfitparams}. 

\begin{table}[tb]
    \centering
    \footnotesize{
    \begin{tabular}{c|c c c c c}
          & $A$  & $\alpha$ & $B$  & $\beta$ & $C$ \\
        \hline
        Nucleus & $(9.7\pm 25.7)\times10^9$\, & $0.77\pm 0.13$\, & $(1.58\pm 0.40)\times 10^4$\, & $-1.44\pm0.05$\, & $0\pm 0.19$ \\
        SuperCDMS & $(1.41\pm 0.16)\times 10^8$ & $0.61\pm 0.006$ & $(3.7\pm 4.1)\times 10^4$ & $-2.7\pm 0.3$ & $0.18\pm 0.01$\\
        Edelweiss & $(1.46\pm 0.28)\times 10^5$ & $0.124\pm 0.003$ & $(1.04\pm 0.55)\times 10^5$ & $-2.6\pm 0.1$ & $0.011\pm 0.002$
    \end{tabular}
    }
    \caption{Best fit values for the parametric model (\ref{nucleusfit}) for the three data sets. The parameters $A,B,C$ are in units of events/(eV g day). }
    \label{bestfitparams}
\end{table}

We use this fit function as the underlying event rate and apply the energy loss as shown in equation (\ref{eq:eloss}), assuming isotropic recoil distribution. In the fit function the exponential component is taken to describe trigger noise and therefore does not reflect true physical recoil events. Thus the energy loss is not applied to this component. We also take the constant component to represent electromagnetic background, resulting in electron recoils and therefore only apply the energy loss to the power law component. However, the results are not very sensitive to this choice.

\begin{figure}[tb]
    \begin{center}
    \includegraphics[width=\linewidth]{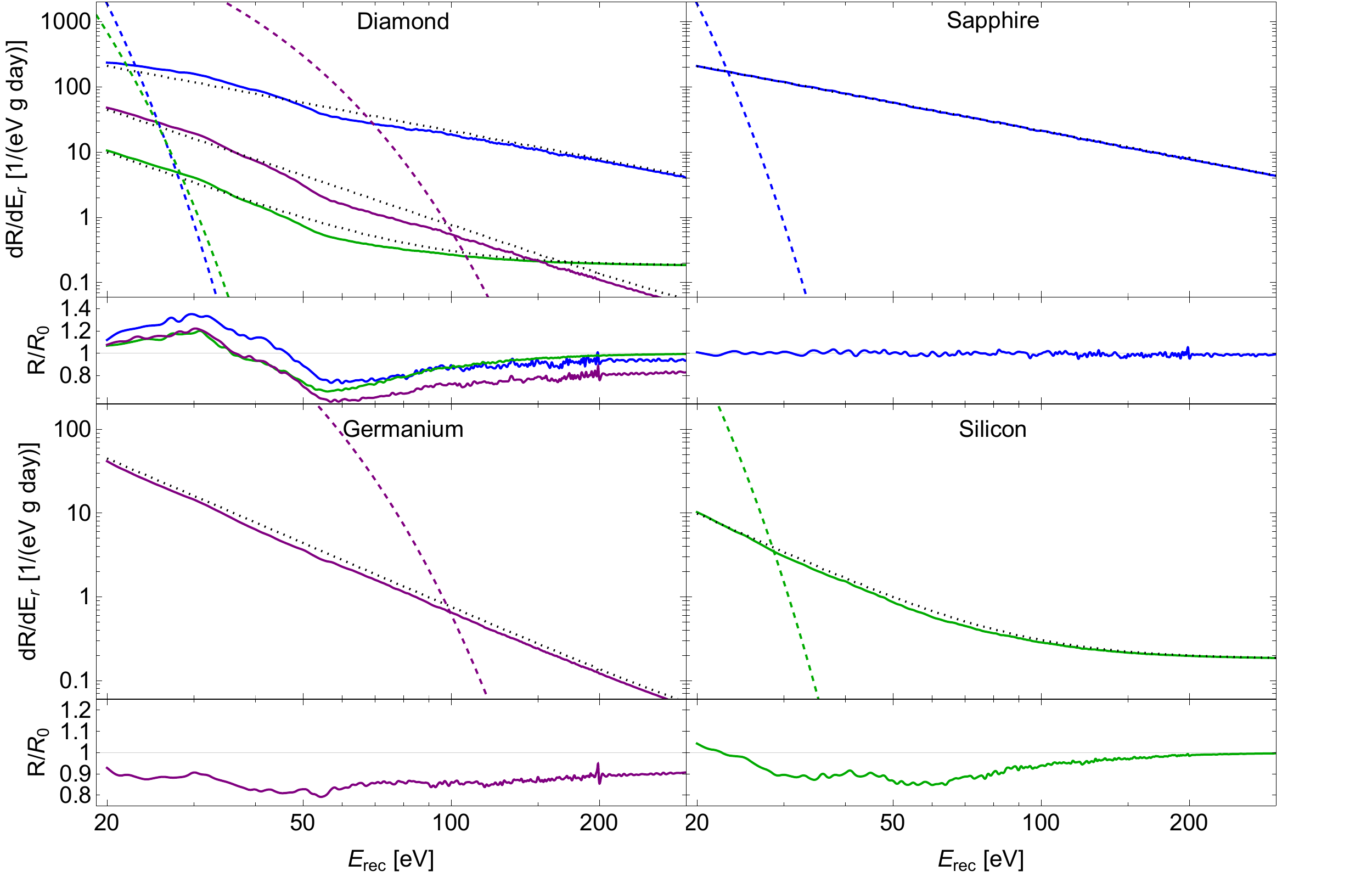}
		\caption{Top left: The observed event rate in diamond for the underlying recoil spectrum given by the power law component of the fit function  (\ref{nucleusfit}). The blue lines correspond to the best fit parameter values for Nucleus data, the green lines for SuperCDMS data and the purple lines for Edelweiss data. The solid curves show the observed rate after the energy loss, and black dotted curves if the energy loss is not simulated. The dashed curves show the exponential component of each fit. The curve in the bottom inset shows the ratio of the rate with/without the energy loss for the power law component. Top right: The same for a sapphire detector, using Nucleus fit. The bottom row shows same results for germanium (left, Edelweiss fit) and silicon (right, SuperCDMS fit). The figure has appeared before in~\cite{Heikinheimo:2021syx}.}
    \label{dRdEplots}
    \end{center}
\end{figure}

The resulting event rates in diamond, sapphire, germanium and silicon are shown in figure \ref{dRdEplots}. As expected, the spectrum in diamond contains a peak around 30 eV, followed by a dip centered at $\sim\! 60$ eV, clearly visible in the bottom inset of the top left panel in the figure, showing the ratio of the event rate with/without the energy loss effect. In sapphire the energy loss effect is insignificant, and for silicon and germanium the threshold is weaker, and located at recoil energies below the range available for our data sets. Therefore there is no peak in the spectrum for these materials, but the energy loss is seen as a slight shift of the spectrum toward lower energy.

To quantify how the energy loss effect in diamond can be used to identify the nuclear recoil origin of the spectrum, we have performed a likelihood ratio test for simulated data sets, using the three best fit functions given in table \ref{bestfitparams} as the physical recoil rate. We obtain the required number of events in a diamond detector in the event selection window [20-200] eV, using 1 eV bins, for a $3\sigma$ observation of the peak/dip feature, as described in more detail in~\cite{Heikinheimo:2021syx}. We find that unless the peak is masked by the exponential noise component, the feature can be observed with just 440 to 710 events in the selection window, depending slightly on the power law index $\beta$ of the fit parameters.

\section{Conclusion}
Nuclear recoils create lattice defects that store part of the recoil energy, resulting in quenching of the observed energy in phonon based detectors. Close to threshold this effect is non-linear and gives rise to a peak in the observed energy spectrum in a diamond detector. We have shown that for a power law spectrum the presence of this peak can be used as an effective test of the nuclear recoil origin of the events.

\paragraph{Funding information}
This work has been supported by the Academy of Finland project $\# 342777$.



\bibliography{IDMproceedings.bib}

\begin{thebibliography}{10}
\providecommand{\url}[1]{\texttt{#1}}
\providecommand{\urlprefix}{URL }
\expandafter\ifx\csname urlstyle\endcsname\relax
  \providecommand{\doi}[1]{doi:\discretionary{}{}{}#1}\else
  \providecommand{\doi}{doi:\discretionary{}{}{}\begingroup
  \urlstyle{rm}\Url}\fi
\providecommand{\eprint}[2][]{\url{#2}}

\bibitem{Strauss:2017cam}
R.~Strauss \emph{et~al.},
\newblock \emph{{Gram-scale cryogenic calorimeters for rare-event searches}},
\newblock Phys. Rev. D \textbf{96}(2), 022009 (2017),
\newblock \doi{10.1103/PhysRevD.96.022009},
\newblock \eprint{1704.04317}.

\bibitem{CPD:2020xvi}
C.~W. Fink \emph{et~al.},
\newblock \emph{{Performance of a large area photon detector for rare event
  search applications}},
\newblock Appl. Phys. Lett. \textbf{118}(2), 022601 (2021),
\newblock \doi{10.1063/5.0032372},
\newblock \eprint{2009.14302}.

\bibitem{Verma:2022tkq}
S.~Verma \emph{et~al.},
\newblock \emph{{Large-mass, low-threshold sapphire detector for rare event
  searches}}  (2022),
\newblock \eprint{2203.15903}.

\bibitem{CRESST:2019axx}
A.~H. Abdelhameed \emph{et~al.},
\newblock \emph{{Description of CRESST-III Data}}  (2019),
\newblock \eprint{1905.07335}.

\bibitem{CRESST:2019jnq}
A.~H. Abdelhameed \emph{et~al.},
\newblock \emph{{First results from the CRESST-III low-mass dark matter
  program}},
\newblock Phys. Rev. D \textbf{100}(10), 102002 (2019),
\newblock \doi{10.1103/PhysRevD.100.102002},
\newblock \eprint{1904.00498}.

\bibitem{DAMIC:2020cut}
A.~Aguilar-Arevalo \emph{et~al.},
\newblock \emph{{Results on low-mass weakly interacting massive particles from
  a 11 kg-day target exposure of DAMIC at SNOLAB}},
\newblock Phys. Rev. Lett. \textbf{125}, 241803 (2020),
\newblock \doi{10.1103/PhysRevLett.125.241803},
\newblock \eprint{2007.15622}.

\bibitem{EDELWEISS:2019vjv}
E.~Armengaud \emph{et~al.},
\newblock \emph{{Searching for low-mass dark matter particles with a massive Ge
  bolometer operated above-ground}},
\newblock Phys. Rev. D \textbf{99}(8), 082003 (2019),
\newblock \doi{10.1103/PhysRevD.99.082003},
\newblock \eprint{1901.03588}.

\bibitem{EDELWEISS:2020fxc}
Q.~Arnaud \emph{et~al.},
\newblock \emph{{First germanium-based constraints on sub-MeV Dark Matter with
  the EDELWEISS experiment}},
\newblock Phys. Rev. Lett. \textbf{125}(14), 141301 (2020),
\newblock \doi{10.1103/PhysRevLett.125.141301},
\newblock \eprint{2003.01046}.

\bibitem{CRESST:2017ues}
G.~Angloher \emph{et~al.},
\newblock \emph{{Results on MeV-scale dark matter from a gram-scale cryogenic
  calorimeter operated above ground}},
\newblock Eur. Phys. J. C \textbf{77}(9), 637 (2017),
\newblock \doi{10.1140/epjc/s10052-017-5223-9},
\newblock \eprint{1707.06749}.

\bibitem{NUCLEUS:2019kxv}
J.~Rothe \emph{et~al.},
\newblock \emph{{NUCLEUS: Exploring Coherent Neutrino-Nucleus Scattering with
  Cryogenic Detectors}},
\newblock J. Low Temp. Phys. \textbf{199}(1-2), 433 (2019),
\newblock \doi{10.1007/s10909-019-02283-7}.

\bibitem{SENSEI:2020dpa}
L.~Barak \emph{et~al.},
\newblock \emph{{SENSEI: Direct-Detection Results on sub-GeV Dark Matter from a
  New Skipper-CCD}},
\newblock Phys. Rev. Lett. \textbf{125}(17), 171802 (2020),
\newblock \doi{10.1103/PhysRevLett.125.171802},
\newblock \eprint{2004.11378}.

\bibitem{SuperCDMS:2018mne}
R.~Agnese \emph{et~al.},
\newblock \emph{{First Dark Matter Constraints from a SuperCDMS Single-Charge
  Sensitive Detector}},
\newblock Phys. Rev. Lett. \textbf{121}(5), 051301 (2018),
\newblock \doi{10.1103/PhysRevLett.121.051301},
\newblock [Erratum: Phys.Rev.Lett. 122, 069901 (2019)],
\newblock \eprint{1804.10697}.

\bibitem{SuperCDMS:2020ymb}
D.~W. Amaral \emph{et~al.},
\newblock \emph{{Constraints on low-mass, relic dark matter candidates from a
  surface-operated SuperCDMS single-charge sensitive detector}},
\newblock Phys. Rev. D \textbf{102}(9), 091101 (2020),
\newblock \doi{10.1103/PhysRevD.102.091101},
\newblock \eprint{2005.14067}.

\bibitem{Proceedings:2022hmu}
A.~Fuss, M.~Kaznacheeva, F.~Reindl and F.~Wagner, eds.,
\newblock \emph{EXCESS workshop: Descriptions of rising low-energy spectra}
  (2022), \eprint{2202.05097}.

\bibitem{Kadribasic:2020pwx}
F.~Kadribasic, N.~Mirabolfathi, K.~Nordlund and F.~Djurabekova,
\newblock \emph{{Crystal Defects: A Portal To Dark Matter Detection}}  (2020),
\newblock \eprint{2002.03525}.

\bibitem{Sassi:2022njl}
S.~Sassi, M.~Heikinheimo, K.~Tuominen, A.~Kuronen, J.~Byggm\"astar, K.~Nordlund
  and N.~Mirabolfathi,
\newblock \emph{{Energy loss in low energy nuclear recoils in dark matter
  detector materials}}  (2022),
\newblock \eprint{2206.06772}.

\bibitem{Kurinsky:2019pgb}
N.~A. Kurinsky, T.~C. Yu, Y.~Hochberg and B.~Cabrera,
\newblock \emph{{Diamond Detectors for Direct Detection of Sub-GeV Dark
  Matter}},
\newblock Phys. Rev. D \textbf{99}(12), 123005 (2019),
\newblock \doi{10.1103/PhysRevD.99.123005},
\newblock \eprint{1901.07569}.

\bibitem{Abdelhameed:2022skh}
A.~H. Abdelhameed \emph{et~al.},
\newblock \emph{{A low-threshold diamond cryogenic detector for sub-GeV Dark
  Matter searches}}  (2022),
\newblock \eprint{2203.11999}.

\bibitem{SuperCDMS:2020aus}
I.~Alkhatib \emph{et~al.},
\newblock \emph{{Light Dark Matter Search with a High-Resolution Athermal
  Phonon Detector Operated Above Ground}},
\newblock Phys. Rev. Lett. \textbf{127}, 061801 (2021),
\newblock \doi{10.1103/PhysRevLett.127.061801},
\newblock \eprint{2007.14289}.

\bibitem{Heikinheimo:2021syx}
M.~Heikinheimo, S.~Sassi, K.~Nordlund, K.~Tuominen and N.~Mirabolfathi,
\newblock \emph{{Identification of the low energy excess in dark matter
  searches with crystal defects}}  (2021),
\newblock \eprint{2112.14495}.

\end{thebibliography}
\bibliographystyle{SciPost_bibstyle}

\nolinenumbers

\end{document}